\documentclass[manuscript]{aastex}
\usepackage{amsmath}

\shorttitle{Oscillatory power in evolving magnetic fields}
\shortauthors{S. Krishna Prasad et al.}
 
\begin{document}
\title{Time-dependent suppression of oscillatory power in evolving solar magnetic fields}             
\author{S. Krishna Prasad} 
\affil{Astrophysics Research Centre, School of Mathematics and Physics, Queen's University Belfast, Belfast, BT7 1NN, UK.}                                  
\email{krishna.prasad@qub.ac.uk}
\author{D. B. Jess}
\affil{Astrophysics Research Centre, School of Mathematics and Physics, Queen's University Belfast, Belfast, BT7 1NN, UK.}
\affil{Department of Physics and Astronomy, California State University Northridge, Northridge, CA 91330, USA.} 
\author{R. Jain} 
\affil{School of Mathematics and Statistics, University of Sheffield, Sheffield S3 7RH (UK)}                                  
\author{P.H. Keys} 
\affil{Astrophysics Research Centre, School of Mathematics and Physics, Queen's University Belfast, Belfast, BT7 1NN, UK.}                                  
\affil{Solar Physics and Space Plasma Research Centre (SP$^2$RC), The University of Sheffield, Hicks Building, Hounsfield Road, Sheffield S3 7RH, UK.}                                  

\begin{abstract}
Oscillation amplitudes are generally smaller within magnetically active regions like sunspots and plage, when compared to their surroundings. Such magnetic features, when viewed in spatially-resolved powermaps, appear as regions of suppressed power due to reductions in the oscillation amplitudes. Employing high spatial- and temporal-resolution observations from the Dunn Solar Telescope (DST) in New Mexico, we study the power suppression in a region of evolving magnetic fields adjacent to a pore. By utilising wavelet analysis, we study for the first time, how the oscillatory properties in this region change as the magnetic field evolves with time. Image sequences taken in the blue continuum, G-band, Ca~\textsc{ii}~K and H$\alpha$ filters were used in this study. It is observed that the suppression found in the chromosphere occupies a relatively larger area confirming previous findings. Also, the suppression is extended to structures directly connected to the magnetic region and is found to get enhanced as the magnetic field strength increased with time. The dependence of the suppression on the magnetic field strength is greater at longer periods and higher formation heights. Furthermore, the dominant periodicity in the chromosphere was found to be anti-correlated with increases in the magnetic field strength.
\end{abstract}

\section{Introduction}
The surface amplitudes of $p$-mode acoustic oscillations have been known to be significantly smaller within magnetically active regions compared to those in the surrounding quiet Sun \citep{1962ApJ...135..474L,1981SoPh...69..233W,1982ApJ...253..386L,1988ESASP.286..315T,1992ApJ...393..782T}. Consequently, magnetic regions are highlighted by locations of reduced power in frequency-filtered powermaps. Several theories have been proposed to explain such power suppression. By studying the power difference of waves travelling inwards and outwards of sunspots, \citet{1987ApJ...319L..27B} found that sunspots absorb as much as 50\% of the incident acoustic waves. It was suggested that the absorbed power could either be scattered or lost due to the mode coupling of magnetoacoustic or pure magnetic waves. More recently, it has been proposed that the absorption mechanism involves partial conversion of $p$-modes into slow magnetoacoustic waves, which are channelled by magnetic fields and lost into the solar interior \citep{1995ApJ...451..372C,2003SoPh..214..201C,2003MNRAS.346..381C}. Also, convection is largely inhibited in locations of strong magnetic field \citep{1952PASP...64...98C} as evidenced in recent sunspot simulations \citep{2006ApJ...641L..73S,2009ApJ...691..640R} hence the excitation of $p$-modes by turbulent convection \citep{1977ApJ...212..243G,1988ApJ...326..462G} is less likely to be efficient in highly magnetic regions. Alternatively, it has been suggested that the acoustic wave functions are modified in the presence of magnetic fields in such a way that the attenuation length of the evanescent $p$-modes is reduced in the solar atmosphere \citep{1996ApJ...464..476J,1997ApJ...476..392H}. This mechanism reduces the observed oscillation amplitudes in magnetic regions without any loss of the actual $p$-mode power. A related effect has been discussed in \citet{2006RSPTA.364..313B} where the authors demonstrate that the existence of wave sources at greater depths in the umbral regions results in reduced oscillatory power at the surface. The possible differences in the line formation heights associated with magnetic and non-magnetic environments which will naturally give rise to different observed amplitudes,  has also been discussed previously \citep{1998ApJ...504.1029H,2002A&A...387.1092J}.

As observations progress to higher cadences, it has been observed that at high frequencies ($>$5 mHz) the power is in fact enhanced surrounding magnetic regions \citep{1992ApJ...394L..65B,1992ApJ...392..739B,2002A&A...387.1092J,2007A&A...471..961M,2013SoPh..287..107R}. Similar behaviour was observed around quiet-Sun network magnetic elements \citep{2001A&A...379.1052K,2010A&A...510A..41K,2012ApJ...744...98C}. At larger heights in the atmosphere, enhanced suppression is found at higher frequencies surrounding active region \citep{2003A&A...401..685M} and quiet-Sun \citep{2001ApJ...554..424J,2007A&A...461L...1V,2010A&A...510A..41K,2013SoPh..282...67G} magnetic fields. Although the exact physical mechanism responsible for these phenomena is not yet clear, currently, the most popular theory that explains the observed features at both low and high frequencies is the interaction of acoustic $p$-mode oscillations with the embedded magnetic fields, which generates fast and slow magnetoacoustic waves in the magnetic canopy through the process of mode conversion \citep{2007A&A...471..961M,2014A&A...567A..62K,2016ApJ...817...45R}.

For this work, our main focus is on the suppression of oscillatory power within magnetic regions at low frequencies. It has been demonstrated by \citet{1998ApJ...504.1029H} and \citet{2002A&A...387.1092J} that low-frequency power suppression increases with magnetic field strength for both Doppler velocity and continuum intensity data. Using multi-height observations, \citet{2007A&A...471..961M} observed that the spatial extent of the suppression region increases with height in the solar atmosphere. More recently, through theoretical modelling work, \citet{2014ApJ...796...72J} predicted that the power deficit in the suppression region also increases with height. Here, we investigate these aspects using our high spatial- and temporal-resolution observations in four different channels obtained simultaneously with the Dunn Solar Telescope (DST). We performed a time-dependent study of the oscillatory power suppression related to a small evolving magnetic concentration that is part of an active region. We present the details of our observations and data preparation in Section~\ref{observ}, the applied analysis methods and results in Section~\ref{anres}, and finally discuss the important conclusions of this work in Section~\ref{concl}.

\section{Observations}
\label{observ}
High-resolution images of active region NOAA~11372 were obtained with the Rapid Oscillations in the Solar Atmosphere \citep[ROSA;][]{2010SoPh..261..363J} and the Hydrogen-Alpha Rapid Dynamics camera \citep[HARDcam;][]{2012ApJ...757..160J} instruments at the Dunn Solar Telescope (DST), New Mexico, on 2011 December 10, for about two hours starting from 17:51~UT. Simultaneous observations were made in four different channels; blue continuum (52{\,}{\AA} bandpass centred at 4170{\,}{\AA}), G-band (9.2{\,}{\AA} bandpass centred at 4305.5{\,}{\AA}), Ca~\textsc{ii} K line core (1{\,}{\AA} bandpass centred at 3933.7{\,}{\AA}), and H$\alpha$ line core (0.25{\,}{\AA} bandpass centred at 6562.8{\,}{\AA}), using HARDcam for H$\alpha$ and three identical ROSA cameras for the other lines. The ROSA continuum and G-band filters correspond to a minimum formation height of 25~km and 100~km \citep{2012ApJ...746..183J}, respectively, whereas the Ca~\textsc{ii} K line core forms below 1300~km \citep{1969SoPh...10...79B} and the H$\alpha$ line core originates near 1500 km \citep{1981ApJS...45..635V} above the photosphere. Note that these heights are only representative as the exact determination of line formation height is quite complex, particularly in the chromosphere \citep[for e.g., see the review by][]{2007ASPC..368...27R}.

 All the images were subjected to speckle reconstruction, amongst other post-processing routines, as detailed in \citet{2014A&A...566A..99K}. The final processed cadences were 2.11~s for the continuum and G-band data, 14.81~s for the Ca~\textsc{ii}~K data, and 1.78~s for the H$\alpha$ data. For each dataset, the counts in each image were normalised to the median of that image to remove any variations in brightness due to changes in the inclination of the Sun during the observations. The seeing conditions remained excellent throughout the duration of observations, except for a brief spell of short-term atmospheric fluctuations that affected a small number of images which were manually identified and replaced through interpolation. A series of calibration images, taken along with the science observations, were used to coalign the data across the different channels. This step brings all of the data to a common plate scale (i.e., that of the blue continuum), which involved interpolating the H$\alpha$ data to match.

The corresponding line-of-sight (LOS) magnetograms, obtained from the Helioseismic and Magnetic Imager \citep[HMI;][]{2012SoPh..275..229S}, onboard the \textit{Solar Dynamics Observatory} \citep[\textit{SDO};][]{2012SoPh..275....3P}, are also used in this study. HMI data were processed following the standard \texttt{hmi\_prep} routine available through {\it{SolarSoft}}. The final pixel scale and cadence of the data were approximately $0{\,}.{\!\!}{\arcsec}6$ and 45~s, respectively. The required coalignment between the HMI and ROSA data was achieved by cross-correlating the ROSA blue continuum and HMI continuum images. This process determined the exact plate scale of the ROSA images to be $0{\,}.{\!\!}{\arcsec}0592$ per pixel.
\begin{figure}
\centering
  \includegraphics[width=15cm]{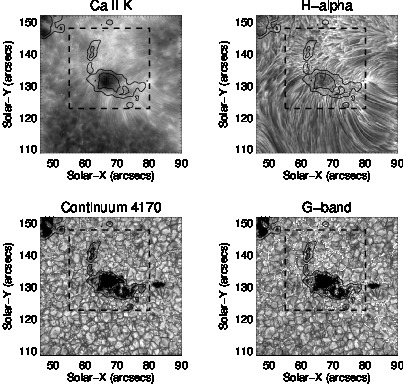}
  \caption{Sample images of a portion of active region NOAA 11372, as observed with the ROSA and HARDcam instruments across four different channels. The overplotted contours, outer and inner, enclose the locations of HMI LOS magnetic fields with strengths $\ge$500~G and $\ge$950~G, respectively. The dashed box highlights the subfield region chosen for creating the powermaps shown in Figure~\ref{fig2}}
  \label{fig1}
\end{figure}
Samples of the images for each channel are shown in Figure~\ref{fig1}. The overplotted contours represent the HMI LOS magnetic fields corresponding to 500~G and 950~G, and highlight the accuracy of the coalignment between the various channels and instruments.

\section{Analysis \& Results} 
\label{anres}
Figure~\ref{fig1} shows sample images relating to a portion of active region AR~11372, as seen in the ROSA and HARDcam channels. The displayed field-of-view encompasses two pores of the same polarity, one close to the centre, and the other at the top-left corner. We consider a subfield region outlined by the dashed box in these images, which covers the central pore. This data is then subjected to wavelet analysis \citep{1998BAMS...79...61T} at each pixel in order to study the spatial distribution of oscillatory power across different period bands. 
\begin{figure}
\centering
  \includegraphics[width=16cm]{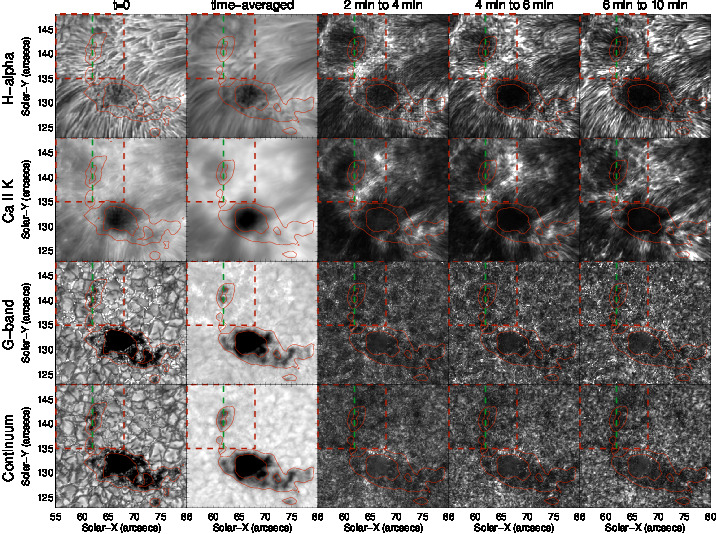}
  \caption{The columns (from left-to-right) show the intensities at the start time, time-averaged intensities and powermaps in three different period bands: 2--4 min, 4--6 min, and 6--10 min, for four different imaging channels as labelled. The overplotted contours represent HMI LOS magnetic field as described in Figure~\ref{fig1}. The dashed red box and the vertical green line mark the locations of subsequent analysis. Darker and brighter regions in the powermaps highlight locations of lower and higher oscillatory power, respectively.}
  \label{fig2}
\end{figure}
Results of the wavelet analysis are shown in Figure~\ref{fig2}. Different rows in this figure correspond to different imaging channels. The first and second columns display actual intensities at the start of the time sequence and those averaged over time, respectively. Powermaps in three different period bands (2--4 min, 4--6 min, and 6--10 min) are displayed in the last three columns, which show the corresponding time-averaged oscillatory power present in these bands. The overplotted contours outline the LOS magnetic field at 500~G and 950~G, at the start of sequence (for the first column) and averaged over time (for the rest of the columns). 

All the powermaps clearly show suppression of oscillatory power, across all bands, at the location of the pore. Interestingly, there is another patch of suppression of similar size visible in the top-left quarter in Ca~\textsc{ii}~K and H$\alpha$ powermaps. This appears to be connected to a fairly small magnetic concentration in this region, as evident from the magnetic field contours. It must be noted that this small magnetic element would easily have been overlooked in visual inspections of the actual magnetograms had it not been for the associated power suppression. In fact, it has been shown that the suppression of surficial acoustic power can be used to probe subsurface magnetic fields well before their emergence \citep{2011SoPh..268..321H}. Thus, to study in detail the observed power suppression and its relation to underlying small magnetic element, we consider an additional subfield outlined by the dashed box in Figure~\ref{fig2}. Inspection of a time-lapse movie of HMI LOS magnetic fields reveals that the smaller magnetic patch in this subfield is constantly evolving from a thin and linear structure at the start of the observations, through to a near circular concentration at the end. One must note that the spatial resolution of HMI is approximately 10 times more coarse compared to ROSA, which can potentially mask other even smaller magnetic concentrations, if present.
\begin{figure}
\centering
  \includegraphics[width=12cm]{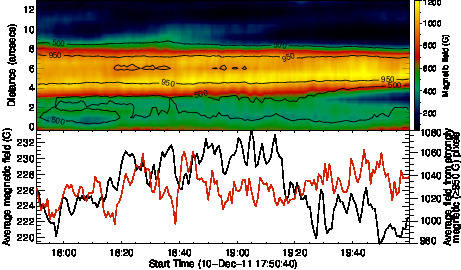}
  \caption{Evolution of the magnetic field within the subfield region outlined by a dashed box in Figure~\ref{fig2}. \textit{Bottom}: The black line represents the average magnetic field over the full subfield, while the red line shows the average computed only from strongly magnetic pixels ($\ge$950~G; inner contours). The axis for the red line is shown on the right. \textit{Top}: Time-distance representation of the magnetic field extracted from the locations marked by the vertical green line in Figure~\ref{fig2}. Distance `0' corresponds to the bottom of the slit, while the applied colour scaling is shown on the right. Also, the temporal location of 500~G and 950~G field strength is highlighted through overplotted contours. An animation showing the spatial distribution of the magnetic field, and its evolution, is available online.}
  \label{fig3}
\end{figure}

The evolution of the average magnetic field in this region is shown in Figure~\ref{fig3}. The black line in the bottom panel represents the average taken over the entire subfield, while the red line represents the average taken only from strongly magnetic pixels, i.e., from locations where the magnetic field $>$950~G. It appears that the strong field average increases with time, even in the later stage when the average for the whole region has decreased. This suggests more of a reorganisation of magnetic flux, rather than the actual emergence of new magnetism. The top panel in this figure shows the time-distance map generated from the location of the vertical green line marked in Figure~\ref{fig2}. Contours outlining locations with magnetic fields $\ge$500~G and $\ge$950~G are also overplotted on this map. It is evident that the vertical extent of the magnetic element is gradually collapsed until the end of the time sequence, where it forms a more concentrated patch. A time-lapse movie of the magnetic field in this region is provided online to clearly show this behaviour. 

\subsection{Time-resolved powermaps}
To understand how the evolving magnetic field in the subfield (outlined by dashed box) in Figure~\ref{fig2} influences the oscillatory power in this region, we extract the time-dependent power from wavelet analysis to construct powermaps at different instants for the same 2--4, 4--6, and 6--10 min period bands.
\begin{figure}
\centering
  \includegraphics[width=15cm]{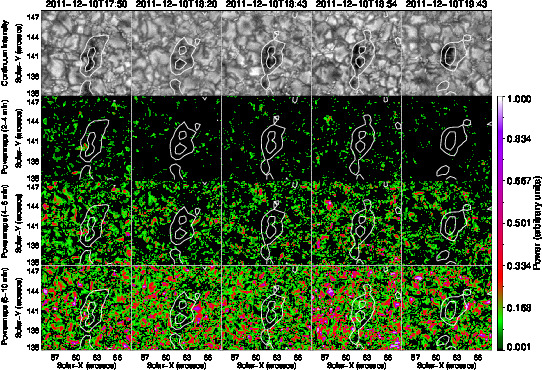}
  \caption{Time-resolved powermaps and the respective intensity images from the continuum channel at five different instants during the observations. Each column displays results for a particular instant in time (shown at the top), while rows from the bottom to top display powermaps in 6--10 min, 4--6 min and 2--4 min, with the top row revealing the corresponding intensity. The contours overplotted denote the locations of magnetic fields at 500~G and 950~G, as obtained from the nearest-time HMI LOS magnetogram. Oscillation power can be seen to be suppressed in the magnetic region. A single colour scale (shown on right) is used for all the period bands, which highlights the decrease in power with time, in addition to revealing lower oscillatory power at shorter periods.}
  \label{fig4}
\end{figure}
\begin{figure}
\centering
  \includegraphics[width=15cm]{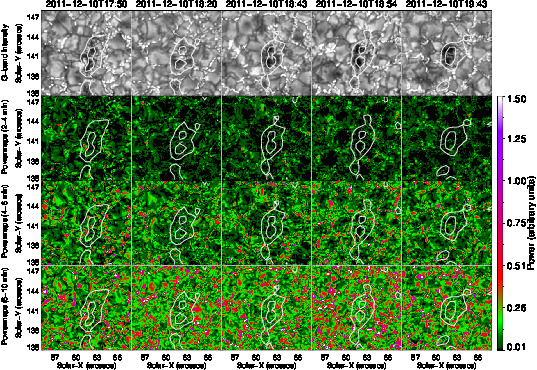}
  \caption{Time-resolved powermaps and the respective intensity images for the G-band channel. Descriptions of the individual panels are the same as those described in Figure~\ref{fig4}. Intensity images highlight G-band bright points at the intergranular lanes. Power suppression properties are similar to those observed in the continuum channel.}
  \label{fig5}
\end{figure}
\begin{figure}
\centering
  \includegraphics[width=15cm]{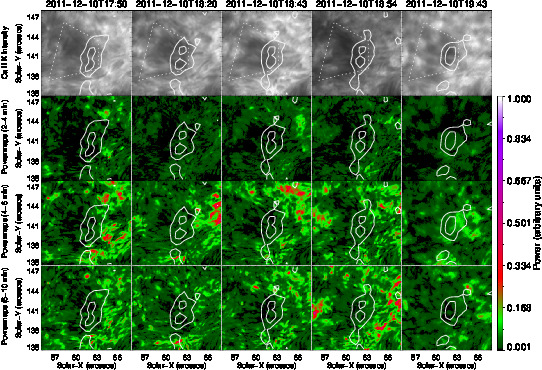}
  \caption{Time-resolved powermaps and the respective intensity images for the Ca~\textsc{ii}~K channel. Descriptions of the individual panels are the same as those described in Figure~\ref{fig4}. The power suppression appears to be spatially extended and asymmetrically distributed towards the left of the magnetic region. The polygon marked over the intensity images roughly encloses the power suppression region used in subsequent analysis.}
  \label{fig6}
\end{figure}
\begin{figure}
\centering
  \includegraphics[width=15cm]{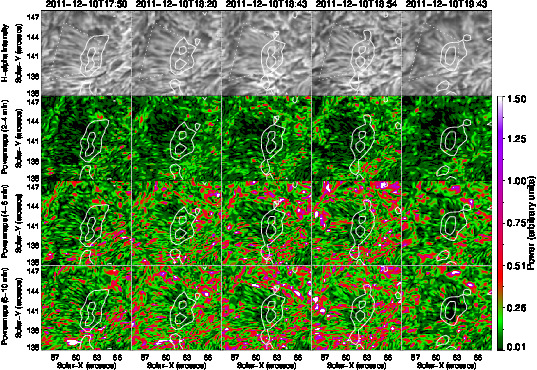}
  \caption{Time-resolved powermaps and the respective intensity images for the H$\alpha$ channel. Descriptions of the individual panels are the same as those described in Figure~\ref{fig4}. Intensity images highlight horizontal canopy structures connected with the magnetic region. Power suppression properties are similar to those observed in the Ca~\textsc{ii}~K channel. The polygon marked over the intensity images roughly encloses the power suppression region used in subsequent analysis.}
  \label{fig7}
\end{figure}
\!\!\!\!\!Figures~\ref{fig4}, \ref{fig5}, \ref{fig6}, \& \ref{fig7} display these time-dependent powermaps at 5 different instants in time for the blue continuum, G-band, Ca~\textsc{ii}~K and H$\alpha$ channels, respectively. In each of these figures, powermaps for discreet bands are shown in the bottom three rows, with the corresponding intensity image displayed in the top row. Respective magnetic field contours, for 500~G and 950~G are also shown using white contours. Each column in these figures represents a time-resolved snapshot of the observations, as specified at the top of each column. The continuum and G-band results (Figures~\ref{fig4} \& \ref{fig5}) show more suppression in the vicinity of the magnetic field as its magnitude becomes stronger and more organised. As already highlighted, one must also consider the difference in the spatial resolutions between the various instruments before trying to make a one-to-one correspondence between the power suppression regions in the ROSA powermaps and the HMI magnetic fields. 

The Ca~\textsc{ii}~K and H$\alpha$ results (Figures~\ref{fig6} \& \ref{fig7}) also show clear evidence for suppression in all period bands at all instants. However, the suppression region appears to be larger and asymmetrically distributed towards the left of the magnetic patch. This appears to be a consequence of the horizontal magnetic canopy structures extending towards the left side, as evident in the H$\alpha$ intensity images. The remarkable structuring visible in the suppression regions of the H$\alpha$ powermaps is also worth noting. The spatial extent of the power suppression increases with height above the photosphere in agreement with the findings of \citet{2007A&A...471..961M}. However, this is the first time it has been observed in higher regions of chromosphere using Ca~\textsc{ii} K and H$\alpha$ images.

\subsection{Time-distance powermaps}
The time-resolved powermaps shown in the previous subsection for 5 different instants of time, revealed the spatial distribution and extent of power at different times spanning the magnetic field evolution. However, to get a more clear picture of the time-dependent nature of the power suppression with the evolving field, we constructed time-distance powermaps. To produce these, we considered a 10-pixel wide slit ($\approx$$0{\,}.{\!\!}{\arcsec}592$ or 430~km) placed on top of the time-resolved powermaps, as shown by the dashed green line marked in Figure~\ref{fig2}, and averaged the power lying across the slit. 
\begin{figure}
\centering
  \includegraphics[width=12cm]{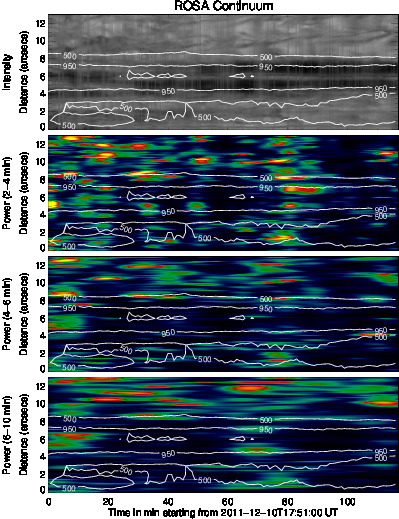}
  \caption{(From top to bottom) Time-distance maps of the continuum intensity and the corresponding power in the three period bands corresponding to 2--4 min, 4--6 min, and 6--10 min. These maps are constructed from the vertical green slit shown in Figure~\ref{fig2}. Distance `0' corresponds to the bottom of the slit. The overplotted contours highlight the locations of magnetic fields at 500~G and 950~G, as labelled. Stronger fields clearly coincide with locations of power suppression.}
  \label{fig8}
\end{figure}
\begin{figure}
\centering
  \includegraphics[width=12cm]{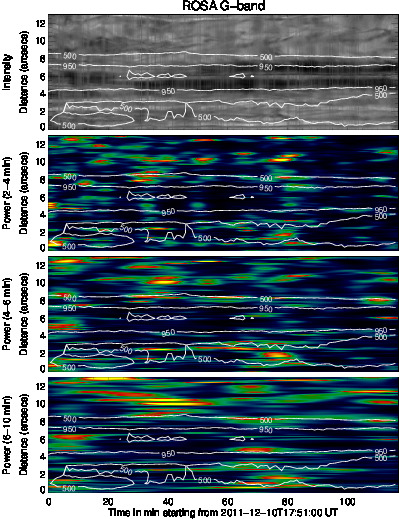}
  \caption{Same as Figure~\ref{fig8}, but for the G-band channel. The strong field regions suppress the oscillatory power in all bands. The bright line at approximately 6$\arcsec$ in the intensity image coincides with the locations of occasional dips in magnetic field (as identified from the magnetic field contour `islands') and shows slight power enhancement in the 6--10 min band.}
  \label{fig9}
\end{figure}
\begin{figure}
\centering
  \includegraphics[width=12cm]{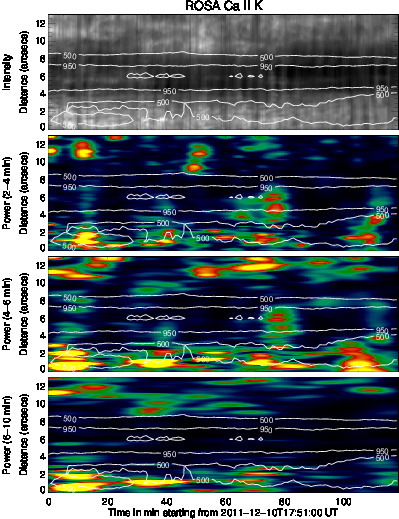}
  \caption{Same as Figure~\ref{fig8}, but for the Ca~\textsc{ii}~K channel. Power suppression again coincides with the locations of strong fields and appears to be more pronounced for longer periodicities. The intensity map clearly shows an oscillatory pattern with $\approx$4 min periodicity.}
  \label{fig10}
\end{figure}
\begin{figure}
\centering
  \includegraphics[width=12cm]{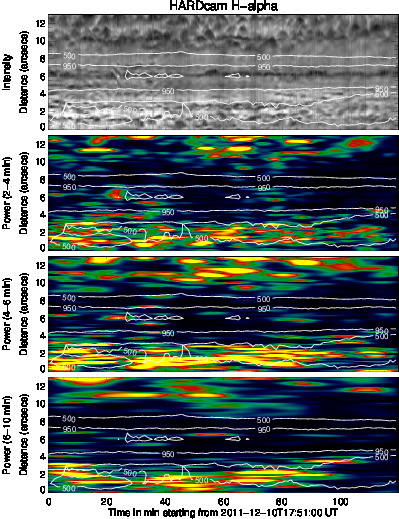}
  \caption{Same as Figure~\ref{fig8}, but for the H$\alpha$ channel. Power suppression properties are similar to those observed in the Ca~\textsc{ii}~K channel. The bright line at 6$\arcsec$ is now replaced with a dark line in intensity, and is accompanied by occasional enhancements in power.}
  \label{fig11}
\end{figure}
\!\!\!\!\!The obtained time-distance powermaps, along with the corresponding intensity plots, are shown in Figures~\ref{fig8}, \ref{fig9}, \ref{fig10}, \& \ref{fig11} for the continuum, G-band, Ca~\textsc{ii}~K and H$\alpha$ channels, respectively. The overplotted contours on these figures enclose the corresponding spatial locations demonstrating field strengths $\ge$500~G and $\ge$950~G as annotated. It is clear from these maps that the region between 4$\arcsec$ -- 7$\arcsec$ distance (from the bottom of the slit) has relatively lower power across all period bands, which coincides with the locations of stronger magnetic fields. The time-distance maps in intensity reveal a brightening (albeit a darkening in H$\alpha$) at the centre of the magnetic patch (at $\approx$6$\arcsec$) where the field strength is slightly weaker at certain times as indicated by the 950~G contour `islands'. The intensity map in Ca~\textsc{ii}~K (Figure~\ref{fig10}) shows a clear oscillatory pattern with a periodicity of $\approx$4~min. A similar periodicity is also found in the HMI magnetic field measurements suggesting they are likely to be magnetoacoustic waves. It is possible that they are driven by $p$-modes originating below the photosphere \citep[see][]{2014ApJ...796...72J, 2015ApJ...812L..15K}. In predominantly photospheric channels, the amplitudes are usually small making them difficult to detect in the continuum and G-band channels and perhaps also the reason why they are not easily discernible in the more chromospheric H$\alpha$ bandpass.

\subsection{Temporal variation of power}
While the time-distance powermaps discussed in the previous subsection reveal the (partly) spatial and temporal behaviour of power in the suppression region, it is not clear whether there is any increase or decrease in power as the magnetic field evolves with time. Hence, we present here how the average power in the suppression region varies with time. To do this, first we interpolate HMI data both spatially and temporally to match the resolution of the ROSA channels. Then we identify the locations of strong magnetic fields (i.e., $\ge$950~G) at each instant and average the power contained within these locations for each period band to generate time evolving power. While scaling the HMI observations to the spatial and temporal resolutions of ROSA involves considerable interpolation, the revised HMI data is solely used to identify magnetic locations over which the ROSA/HARDcam power is averaged. Furthermore, we also verified our results by compressing the ROSA/HARDcam channels spatially and temporally to match the HMI resolution, which resulted in consistent outputs. Also, the power suppression regions in the Ca~\textsc{ii}~K and H$\alpha$ channels do not exactly coincide with the magnetic region as defined by HMI. Therefore, for these channels we selected a separate region bounded by the polygons marked in Figures~\ref{fig6} and \ref{fig7}, which roughly encloses the magnetic patch and the horizontal structures connected to it, and obtained the average power within that isolated region.
\begin{figure}
\centering
  \includegraphics[width=12cm]{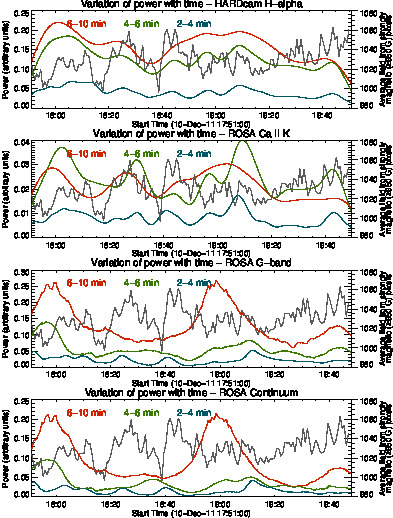}
  \caption{Variations in the average power for different period bands as a function of time. The four panels display results for the four channels studied. In each panel, red, green and blue lines denote the time evolution of power for the different period bands as annotated. The black line represents the corresponding magnetic field evolution, with its axis shown on right. For the continuum and G-band channels, the power is averaged only from the strong magnetic field locations ($\ge$950~G), whereas for the Ca~\textsc{ii}~K and H$\alpha$ channels, the power is averaged over the polygon marked in Figures~\ref{fig6} and \ref{fig7}. }
  \label{fig12}
\end{figure}
The results are shown in Figure~\ref{fig12}, where different colours represent the different period bands as annotated. The black curve represents the corresponding variation in the photospheric (HMI) magnetic field strength from the original, uninterpolated data. It appears that in general the power across all period bands decreases with increases in the magnetic field (i.e., is anti-correlated). However, the power also appears to fluctuate with time, and naturally one would wish to check whether these small-scale fluctuations are also anti-correlated with changes in magnetic field. But this is difficult to accomplish since the power, which is proportional to the oscillation amplitude, also fluctuates in time due to closely spaced oscillation frequencies \citep{2015ApJ...812L..15K}. A closer inspection of the 4~min oscillations seen in Ca~\textsc{ii}~K indeed shows periodic variations in the amplitude, which resembles a beat pattern. To study the exact quantitative dependence of oscillation power on magnetic field, we construct co-spatial and co-temporal power maps with HMI magnetic field, which are discussed in the following subsection.

\subsection{Quantitative dependence of oscillation power on magnetic field}
\label{mag_pow}
So far we have shown that the oscillation power is suppressed in the presence of a magnetic field, and that the suppression becomes increasingly enhanced as the magnetic field strength increases. In this section, we study the exact quantitative dependence of oscillation power, across different period bands, on the magnetic field. To perform this, we compressed all of the time-resolved powermaps, both spatially and temporally, to construct co-spatial and co-temporal powermaps with HMI magnetic field. This allowed us to make a one-to-one mapping of the oscillation power with the magnetic field. We then considered bins of 100~G and estimated the means and standard deviations of oscillatory power contained within each period band, for each bin.
\begin{figure}
\centering
  \includegraphics[width=12cm]{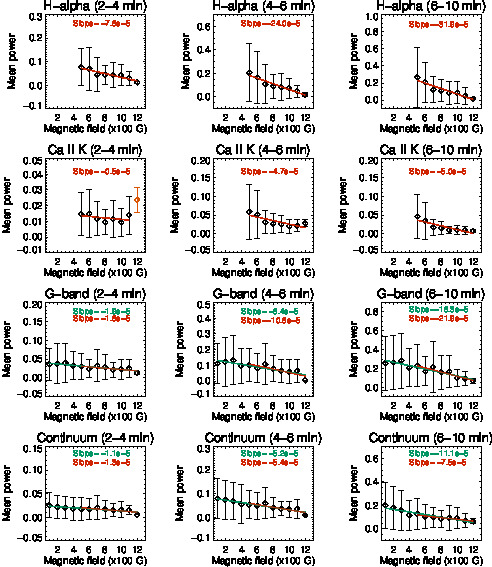}
  \caption{Dependence of the oscillatory power on the magnetic field strength across different channels and period bands as annotated. In each plot, diamonds represent the average power within a 100~G bin starting from that value, and the error bars denote the respective standard deviations. The overplotted sold lines are linear fits to the data. Locations with field strengths above 100~G are considered for the continuum and G-band channels, while fields above 500~G are considered for the Ca~\textsc{ii}~K and H$\alpha$ channels. Consequently, two different fits are performed for the photospheric channels (i.e., continuum and G-band; see text for details). The last point in the Ca~\textsc{ii}~K plot for the 2--4 min band (shown in red) is discarded while making the fit. The obtained gradient values are listed in each plot. Oscillation power clearly decreases with increasing magnetic field strength.}
  \label{fig13}
\end{figure}
The obtained mean values are plotted as diamonds in Figure~\ref{fig13} for each channel, with the corresponding standard deviations used to create the error bars. For the continuum and G-band channels, only those locations with a minimum magnetic field strength of 100~G are chosen. For Ca~\textsc{ii}~K and H$\alpha$, this limit is set at 500~G. The reason a higher limit is imposed for these channels is that a large portion of the structures in the subfield appear to be connected to the magnetic region (see Figure~\ref{fig7}) as previously described, which makes it less meaningful to map one-to-one across the full subfield. Clearly, all the plots in Figure~\ref{fig13} show an enhanced suppression of power with increasing magnetic field strength. Here, we would like to emphasise that the error bars do not represent actual errors, but are instead the standard deviations for each value within the corresponding bin. Consequently, the extent of the error bars is smaller at higher field strengths as a result of more limited statistics. 

In order to quantitatively compare suppression across different channels and period bands, the mean power values were fit linearly as shown in Figure~\ref{fig13}. For the continuum and G-band data, separate fits were made for the full range of values (green line), and for values corresponding to magnetic field strengths above 500~G (red line). The last point in the Ca~\textsc{ii}~K plot for the 2--4 min band (marked in red) was ignored as an outlier while making the fit. This is justified considering the poor statistics associated with the highest field strengths. The obtained gradients are listed in each plot in the Figure. Comparing the gradient values across different period bands, it appears that the suppression is steeper for longer periods. A similar comparison across different channels, e.g., the continuum, G-band, and H$\alpha$, suggests that the suppression is also steeper at higher formation heights. The different behaviour observed in the Ca~\textsc{ii}~K plots is likely a result of its wide passband ($\approx$1{\,}$\AA$) that naturally covers an extended range of solar atmospheric heights. In fact, \citet{2009A&A...500.1239R} has shown that the contributions from bright photospheric wings swamp the dark chromospheric structures in Ca~\textsc{ii} K filtergrams, even for a bandpass narrower than that used in the present study.

\subsection{Dominant period}
In this section we discuss the effects of evolving magnetic fields on the dominant periods and the corresponding spatial distributions. We identify peaks in the wavelet power spectra at each pixel and consider the period at the strongest peak (with highest power) as the dominant period. Since the wavelet analysis gives us time-resolved power spectra, we were able to construct {\it{time evolving}} dominant period maps. Power spectra with periodicities only up to 10 minutes are considered in this analysis.
\begin{figure}
\centering
  \includegraphics[width=15cm]{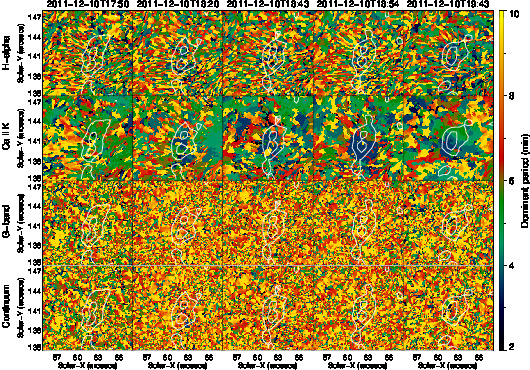}
  \caption{Dominant period maps (maps showing the period with highest oscillatory power at each spatial location) in all four imaging channels at five different instants during the observations as annotated. Power spectra only in the 0--10 min range is considered to generate these plots. The color scale is shown on the right, with the corresponding magnetic field contours overplotted similar to those displayed in Figures~\ref{fig4} -- \ref{fig7}. Ca~\textsc{ii}~K and H$\alpha$ maps clearly show a drift toward shorter periods as the field strength increases with time.}
  \label{fig14}
\end{figure}
The generated dominant period maps at 5 different instants during the observation period are shown in Figure~\ref{fig14} for all four channels. The white contours overplotted correspond to a magnetic field strength of 500~G and 950~G. The continuum and G-band maps are dominated by red and yellow patches synonymous with longer ($>$6~min) periods. Although sparsely populated, the green and blue patches in these maps, corresponding to 5~min and lower periods, are predominantly in the intergranular lanes (see Figures~\ref{fig4} and \ref{fig5}). However, the major magnetic concentration in this region does not seem to have any influence on the distribution of dominant periods seen in these maps. On the other hand, the Ca~\textsc{ii}~K and H$\alpha$ maps show preferentially shorter periods at the magnetic patch, alongside longer periods in the horizontal structures connected to it. The increasing green and blue colours, found in the vicinity of the magnetic patch from the start of the time sequence to the end, indicates the dominance of shorter periods with increasing field strength. The changing spatial distribution of the surrounding red/yellow patches in Ca~\textsc{ii}~K is also worth noting. 
\begin{figure}
\centering
  \includegraphics[width=12cm]{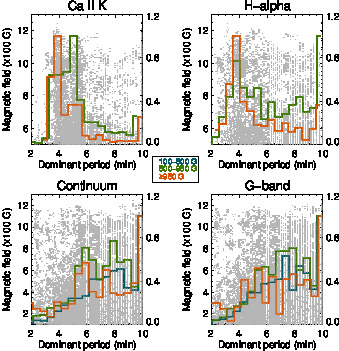}
  \caption{Scatter plots showing the dependence of the dominant period on the magnetic field strength for all four imaging channels. Locations with field strengths above 100~G are considered for the continuum and G-band channels, and those above 500~G are considered for the Ca~\textsc{ii}~K and H$\alpha$ channels. The overplotted solid lines show histograms for this data within different magnetic field ranges (listed in the legend at the centre) as identified by their color. All the histograms are normalised to their respective maxima before plotting. The axis for the histograms is displayed on the right.}
  \label{fig15}
\end{figure}

Figure~\ref{fig15} shows a scatter plot of the dominant periods versus the magnetic field strengths for all four channels. All of the ROSA/HARDcam data has been compressed to match the HMI resolution, and is therefore similar to that described in section~\ref{mag_pow}. Again, only pixel locations corresponding to magnetic fields $>$100~G were considered for the continuum and G-band channels, and those corresponding to $>$500~G were included for the Ca~\textsc{ii}~K and H$\alpha$ channels. The overplotted solid lines in colour show histograms of the dominant periods for the magnetic field ranges listed in the figure legend. Each of the histograms were normalised to their respective maxima before plotting. While the scatter plots indicate all of the possible dominant periods throughout the magnetic field range, the histograms suggest many of the pixel locations have dominant period $>$5 min in the continuum and G-band channels, irrespective of the strength of the field. A similar inference can also be drawn from the dominant period maps. However, it is important to note that the scatter plots include data from all temporal locations. Plots for Ca~\textsc{ii}~K and H$\alpha$ show a majority of the pixel locations with stronger fields ($>$950~G) have a dominant period at $\approx$4 min. This value is shifted to about 5 min for pixel locations with relatively weaker fields (500~G -- 950~G) in Ca~\textsc{ii}~K. For H$\alpha$, the dominant period appears to have shifted to about 10 min for weaker field locations, but still a significant fraction of them show dominant power at $\approx$4~min. The difference in behaviour of Ca~\textsc{ii}~K and H$\alpha$ channels in the weak field regime, i.e., at the locations between the outer and the inner contours shown in Figure~\ref{fig14}, can also be understood to have arisen from their distinct passbands. The 1{\,}{\AA} wide Ca~\textsc{ii}~K channel can have contributions from slightly inclined fields in the lower atmosphere, shifting the dominant period from 4~min to 5~min, while the narrowband H$\alpha$ channel has contributions mainly from the nearly horizontal structures (see Figure~\ref{fig7}), which perhaps can raise the dominant period to $\approx$10~min. 

\section{Conclusions}
\label{concl}
Observations of active region NOAA~11372 reveal a reduction in $p$-mode oscillatory power (suppression) adjacent to a magnetic pore. The corresponding HMI LOS magnetic fields disclose a small magnetic concentration within this region, which is found to evolve with time. During the observation period, the magnetic element appears to change from a thin linear shape into a more uniform circular shape. The overall magnetic field strength for this element is also found to increase with time. We studied the effect of the evolving field on the time-dependent oscillatory power through the application of wavelet analysis. The power suppression regions in the continuum and G-band images are found to coincide with the locations of the magnetic element, whereas the power suppression found in Ca~\textsc{ii}~K and H$\alpha$ images is spatially extended and asymmetrically distributed towards the left of the magnetic element. The asymmetric distribution of horizontal canopy structures in the chromosphere, which are connected to the magnetic region, most likely explains the different behaviour in the latter channels. Power suppression is found to increase with time as the magnetic field is increased. Detailed analysis on the dependence of the oscillatory power on the magnetic field strengths revealed negative gradients, suggesting this is indeed true. A quantitative comparison of this dependence between different period bands, and across different channels, indicates that the power suppression is steeper for longer periods and at higher atmospheric heights. 

Using a theoretical model, \citet{2014ApJ...796...72J} showed that the power deficit in the suppression region increases with height above the photosphere. They also found that this behaviour manifests with a larger effect when the magnetic field is stronger, which explains the steeper dependence on magnetic field we found at larger heights. At the locations of higher field strengths, which were closer to the centre of the magnetic element (see Figure~\ref{fig3}), the magnetic fields are likely to be less inclined than those in the  surroundings with relatively lower field strengths. This difference in inclination can cause more effective suppression in longer periods at higher field strengths (for instance, a recent study by \citet{2013ApJ...779..168J} demonstrates the effect of magnetic field inclination on the dominant oscillation period observed) and thus explains the steeper dependence of the oscillation power on the magnetic field for longer periods. We have not computed power suppression ratios to study the frequency dependence shown by \citet{2014ApJ...796...72J} and \citet{1998ApJ...504.1029H} since our field of view mainly encompasses the active region and therefore does not provide an inherently quiet region to make them statistically meaningful.

In addition, we studied the spatial distribution of the dominant period, and its evolution, with time. The dominant period extracted from the continuum and G-band intensities did not show any influence from the magnetic concentrations, whereas the Ca~\textsc{ii}~K and H$\alpha$ time series displayed shorter periods in the vicinity of stronger magnetic fields. Here, the dominant periodicity was also found to shorten with time as the magnetic field strength increased. Again, this behaviour may be related to a decrease in the inclination of the field as the field strength increases.

The observed increase in the spatial extent of the suppression region with height is in agreement with the results of \citet{2007A&A...471..961M} and \citet{2014ApJ...796...72J}. \citet{2007A&A...471..961M} explained their observations based on a mode-conversion model where the power suppression is primarily due to the partial conversion of $p$-modes into magnetoacoustic waves that go undetected \citep{2003MNRAS.346..381C}. The model by \citet{2014ApJ...796...72J} also involves the generation of magnetoacoustic waves via the buffeting of p-modes, but here the power suppression is mainly a result of the difference in the evanescence of compressive waves within and outside the magnetic flux tube, producing shorter attenuation lengths inside the tube. In addition, each magnetic element is considered as a collection of thin magnetic fibrils where no radial structuring is present. Despite the differences, both models could reproduce the increasing spatial extent of the suppression region with atmospheric height. Therefore, the exact physical explanation behind these phenomena remains open and requires further theoretical and observational studies. In the near future we hope to utilize coordinated multi-height observations with spectropolarimetric sensitivites combined with realistic atmospheric models to address important issues such as the location of equipartition ($\beta=$1) layer with respect to the observed spectral lines, identifying the slow/fast nature of the observed oscillations, etc., that might help resolve this problem.

\acknowledgements
The authors thank the referee for useful comments. D.B.J. thanks STFC for an Ernest Rutherford Fellowship in addition to a dedicated standard grant that allowed this project to be undertaken. The HMI data used here are courtesy of NASA/SDO and HMI science teams.

{\it Facilities:} \facility{Dunn (ROSA, HARDcam)}, \facility{SDO (HMI)}.

%\bibliographystyle{apj.bst}
%\bibliography{kpref}

\end{document}